\documentclass[apj]{emulateapj}
\usepackage{latexsym}
\usepackage{graphicx}
\usepackage{amssymb}
\usepackage{longtable}
\usepackage{apjfonts}
\usepackage{mathptmx}
\usepackage{lscape}

\def\lapp{\ifmmode\stackrel{<}{_{\sim}}\else$\stackrel{<}{_{\sim}}$\fi}
\def\gapp{\ifmmode\stackrel{>}{_{\sim}}\else$\stackrel{<}{_{\sim}}$\fi}

\shorttitle{Timing of millisecond pulsars in NGC\,6752 - II}
\shortauthors{Corongiu et al.}

\begin{document}


\title{Timing of millisecond pulsars in NGC\,6752 - II. \\
  Proper motions of the pulsars in the cluster outskirts}


\author{A. Corongiu\altaffilmark{1,2},
  A. Possenti\altaffilmark{1}, A.G. Lyne\altaffilmark{3},
  R.N. Manchester\altaffilmark{4}, F. Camilo\altaffilmark{5},
  N. D'Amico\altaffilmark{1,2} and J.M. Sarkissian\altaffilmark{6}}

%
%

\altaffiltext{1}{INAF - Osservatorio Astronomico di Cagliari, 
              Loc. Poggio dei Pini, Strada 54, 09012 Capoterra (CA), Italy}
\altaffiltext{2}{Universit\`{a} degli Studi di Cagliari, Dip. di Fisica, 
	      S.S. Monserrato-Sestu km 0,700, 09042 Monserrato (CA), Italy}
\altaffiltext{3}{University of Manchester, Jodrell Bank Observatory,
              Jodrell Bank, Macclesfield, Cheshire, SK11 9DL, UK}
\altaffiltext{4}{Australia Telescope National Facility, Commonwealth
              Scientific and Industrial Research Organization,
              P.O. Box 76, Epping, NSW 1710, Australia}
\altaffiltext{5}{Columbia Astrophysics Laboratory, Columbia University,
              550 West, 120th Street, New York, NY 10027}
\altaffiltext{6}{Australia Telescope National Facility, CSIRO, Parkes
              Observatory, P.O. Box 276, Parkes, New South Wales 2870,
              Australia}


\begin{abstract}
Exploiting a five-year span of data, we present improved timing
solutions for the five millisecond pulsars known in the globular
cluster NGC\,6752. They include proper motion determinations for
the two outermost pulsars in the cluster, PSR\,J1910-5959A and
PSR\,J1910-5959C. The values of the proper motions are in agreement
with each other within current uncertainties, but do not match (at
4$\sigma$ and 2$\sigma$ level respectively) with the value of the
proper motion of the entire globular cluster derived in the optical
band.  Implications of these results for the cluster membership of the
two pulsars are investigated. Prospects for the detection of the
Shapiro delay in the binary system J1910-5959A are also discussed.
\end{abstract}



\section{Introduction}

The globular cluster (GC) NGC\,6752 is known to host five millisecond
pulsars (MSPs) \citep[][hereafter Paper\,I]{dlm+01,dpf+02}.  PSR
J1910$-$5959B and PSR\,J1910$-$5959E (hereafter PSR\,B and E
respectively) reside in the central region of the GC and show large
negative $\dot{P}$ values, which are interpreted as an effect of the
GC gravitational potential well (Paper\,I). This in turn implies a
large mass-to-light ratio in the central region of NGC\,6752.
\citet{fps+03} recalculated the center of gravity and studied the
luminosity profile of this cluster: combining their HST data with the
$\dot{P}$ value of PSR\,B and PSR\,E, they put a firm lower limit on
the central mass-to-light ratio of $M/L_{V}\,\gapp\,5.5\,{\rm
M_\odot}/{\rm L_\odot}$. Also PSR\,J1910$-$5959D (PSR\,D) is located
close to the GC center. Its $\dot{P}$ value is positive and of the
same order of magnitude of PSR\,B and E, suggesting that also for
PSR\,D the $\dot{P}$ value is dominated by the gravitational potential
well (Paper\,I).  PSR\,J1910$-$5959C\footnote{Note that, to conform
with currently accepted practice, all pulsars associated with the cluster have
been given a J2000 name with the same rounded coordinates,
corresponding approximately to the cluster centre.} (PSR\,C) is
located at a projected distance $\theta_{\perp}\,=\,2\farcm6$ from the
GC center (Paper\,I), which is much larger than the cluster's core
radius $r_c\,=\,5\farcs2\pm 2\farcs4$ \citep{fps+03}. The binary
pulsar J1910$-$5959A (PSR\,A) is located at an even larger distance
from the GC center ($\theta_{\perp}\,=\,6\farcm4$, Paper\,I),
the largest offset known for a GC pulsar.

The positions of PSR\,A and PSR\,C are unexpected since mass
segregation should have driven the two neutron stars close to the GC
center in a time scale ($\lapp$\,1\,Gyr) much shorter than the time
since their formation ($\sim$\,10\,Gyr). In particular,
\citet{cpg02,cmp03} explored various scenarios to explain the unusual
position of PSR\,A, invoking a dynamical encounter in the inner region
of the GC. The most probable picture is that PSR\,A was originally in
the GC central regions and it has been expelled to the outskirts by
the interaction with either a single massive black hole (BH) or a
binary [BH\,+\,BH] of unequal mass. Timing results in
Paper\,I indicated a low-mass white dwarf as the most probable
companion for PSR\,A. This has been confirmed by \citet{bvkh03} and
\citet{fpsn03}, who identified with Hubble Space Telescope
observations the companion of PSR\,A with a helium white dwarf star
whose mass is $M_{\it co}\simeq\,0.17-0.20\,{\rm M_\odot}$ and whose
photometric properties are compatible with its belonging to NGC\,6752.

The issue of the association of PSR\,A to NGC\,6752 has been recently
revisited using spectroscopic observations of the optical companion to
the pulsar, performed with the ESO-VLT.  \citet{cfpd06} found full
agreement (at $1\sigma$) between the radial velocity of the center of
mass of the binary $\gamma=-28.1\pm4.9$ km s$^{-1}$ and the overall
cluster radial velocity $v_{\rm 6752}=-27.9\pm0.8$ km s$^{-1}$,
obtained by \citet{har96} (catalog revision 2003) averaging various
determinations.  This is a strong indication in favour of the
association of the pulsar with NGC\,6752. However, using the same data
set, \citet{bkkv06} compared the systemic velocity of the binary with
that of nearby stars which certainly belong to the cluster and
concluded that they are only marginally consistent at $2\sigma$
level.

In this paper we present timing results based on more than five
years of regular observations. In particular,
with a much longer available data span we have been able to measure
proper motions of PSR\,A and PSR\,C. The new timing solutions as well
as the pulse profiles for all the millisecond pulsars are presented in
\S\,2, \S\,3 reports on the proper motion determinations and
implications for the cluster membership of the two pulsars are 
discussed in \S\,4.

\section{Observations and improved timing parameters}

Regular pulsar timing observations of NGC\,6752 have been carried out
since September 2000 with the Parkes 64\,m radio telescope at a
central frequency of 1390\,MHz, using the central beam of the
multibeam receiver or the H-OH receiver.  The hardware system is the
same as that used in the discovery observations \citep{dlm+01}.  The
effects of interstellar dispersion are minimised by using a filterbank
having 512$\times$0.5\,MHz frequency channels for each polarization.
After detection, the signals from individual channels are added in
polarization pairs, integrated, 1 bit-digitized every 125\,$\mu$s
(80\,$\mu$s in recent observations), and recorded to magnetic tape
for off-line analysis. Pulse times of arrival (TOAs) are determined by
fitting a template profile to the observed mean pulse profiles and
analysed using the program {\tt TEMPO}\footnote{see
http://www.atnf.csiro.au/research/pulsar/timing/tempo} and the DE405
solar system ephemeris.

\begin{figure}
\plotone{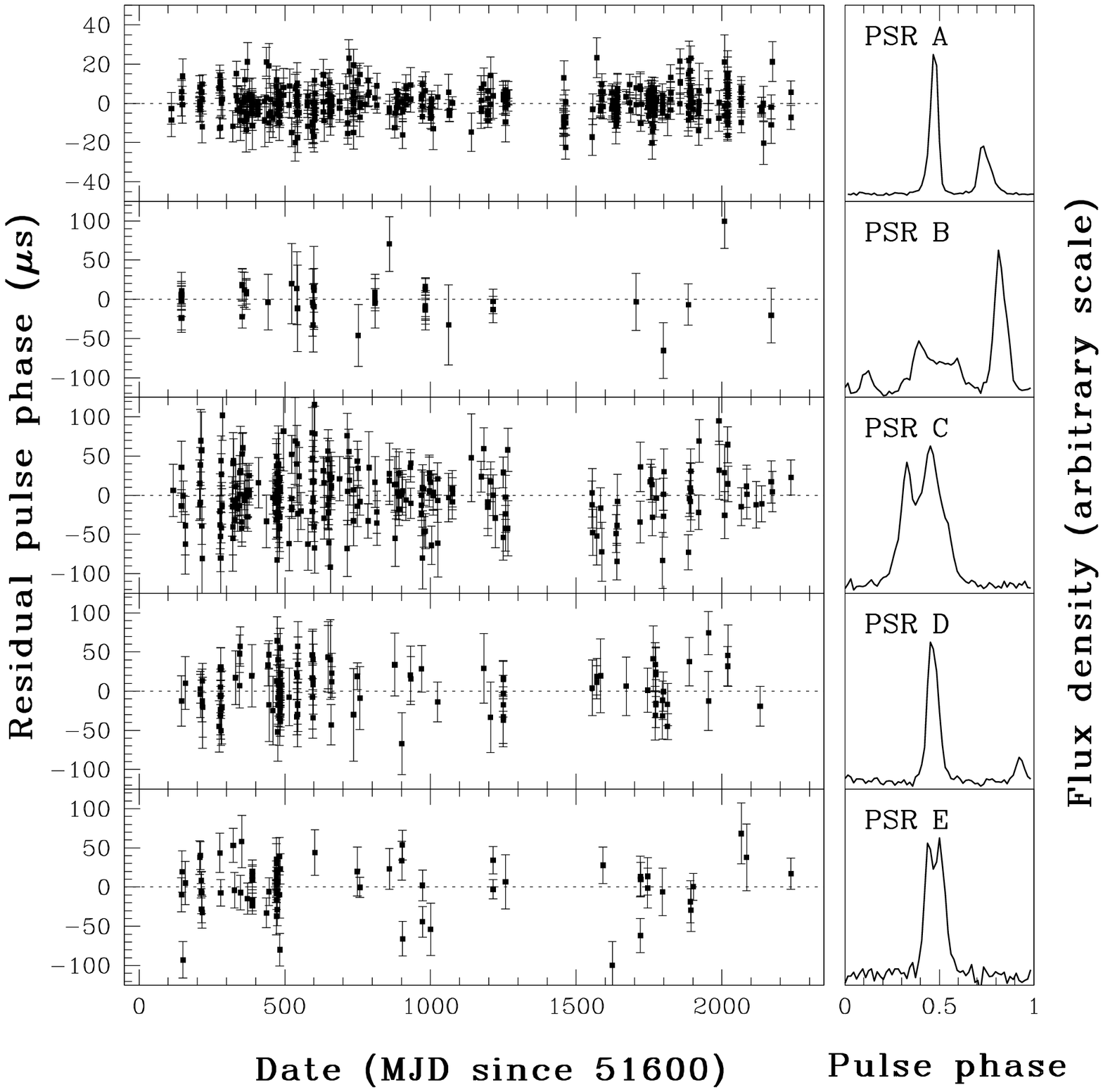}
\caption{Fit residuals ({\em left panels}) and pulse profiles ({\em
right panels}) for the five pulsars known in NGC\,6752. Mean pulse
profiles shown in the right panels are the sum of the observed
profiles with the highest signal-to-noise ratio. The adopted binning
(64 bins) matches the time resolution of the profiles.
\label{fig:profile+residual}}
\end{figure}

Table\,\ref{TabTim} summarizes the best fit values and uncertainties
(chosen to be twice the nominal {\tt TEMPO} errors) for the parameters
entering our timing solutions, whose residuals are displayed in
Figure\,\ref{fig:profile+residual}. The same figure presents a high
signal-to-noise profile obtained for each of the pulsars by folding
the best available data according to the reported ephemerides.

\begin{figure}
\centering
\plotone{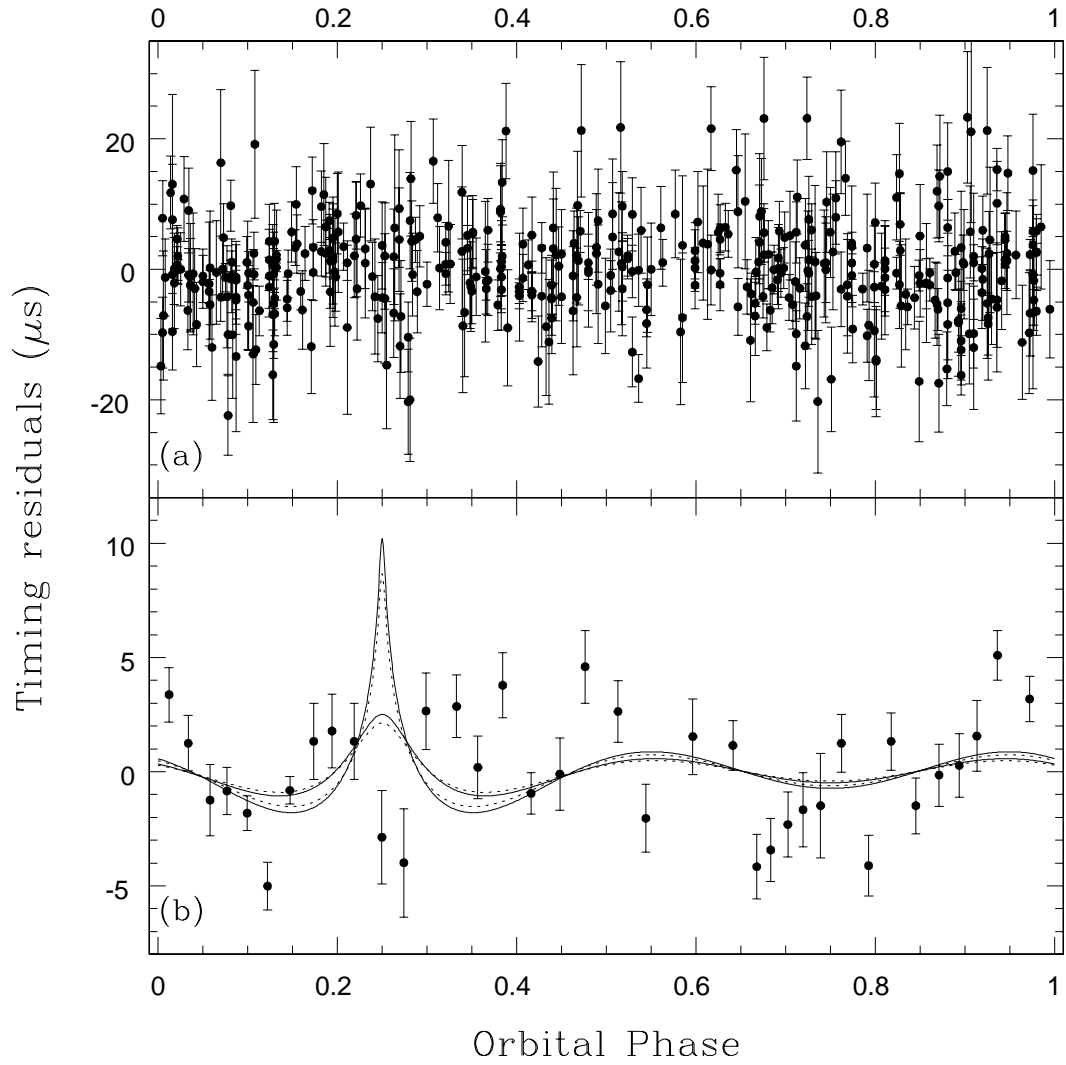}
\caption{{\it (a)} Fit residuals versus orbital phase for PSR\,A
obtained from the timing solution of Table \ref{TabTim}.  {\it (b)}
Timing residuals binned in 42 orbital bins. The central values and the
plotted uncertainties result from a weighted average (and error
propagation) performed on all the available TOAs in each orbital bin.
The lines represent the expected trends of the timing residuals when
the Shapiro delay is not included in the timing model. The two upper
curves are for an orbital inclination $i=89\degr,$ whereas the two
lower curves are for $i=80\degr.$ The mass of the companion star
is taken as $0.20~{\rm M_\odot}$ (solid line) and $0.17~{\rm
M_\odot}$ (dotted line) in both sets of curves.}
\label{fig:orbitres}
\end{figure}

The new positional and rotational parameters at the reference epoch
are all compatibile with those reported in Paper\,I (assuming
$3\sigma$ uncertainties for the values quoted in Paper\,I). However,
the MJD range of the available TOAs is now $\sim 3.5$ times longer
than for Paper\,I and hence the accuracy of the solutions has
improved correspondingly. Orbital parameters for PSR\,A, obtained
using the {\sc ell1} model of {\tt TEMPO}, have also been measured
with a higher precision than in Paper\,I. Figure \ref{fig:orbitres}
shows that no trend is evident in the timing residuals plotted with
respect to the orbital phase for the timing solution given in
Table~\ref{TabTim}.  An additional constraint on the orbit of PSR\,A
has resulted from the recent optical observations of the pulsar
companion.  Spectroscopy \citep{cfpd06,bkkv06} has provided us with a
measurement of the mass ratio in the binary, whereas multi-color
photometry has set the possible range for the mass of the secondary
star \citep{fpsn03,bkkv06}. Combining these results gives a limit on
the orbital inclination $i\gapp 70\degr$ \citep{cfpd06,bkkv06}.

The size of the expected Shapiro delay is nominally larger than the
rms residual of the timing solution (see Table~\ref{TabTim}) for any
$i\gapp70\degr,$ but, except for inclination angles near $90\degr$, a
large part of the Shapiro delay is absorbed in the Roehmer delay
\citep{lcw+01}.  In fact, no clear trend is visible in the timing
residuals even after binning the TOAs in orbital phase (see Figure
\ref{fig:orbitres}), indicating that the magnitude of the unabsorbed
component of Shapiro delay is below the present uncertainty in the
TOAs. Therefore it is not surprising that fitting the available TOAs
with {\tt TEMPO} has not led to any significant determination of the
Shapiro parameter $s$. Inspection of Figure \ref{fig:orbitres} also
shows that the present uncertainties on the TOAs allow us only to
exclude very extreme orbital inclinations $i\gapp 89\degr.$
Simulations show that a factor $\sim 2-3$ improvement in timing
precision is needed in order to obtain a useful constraint on $s$.
This will require an additional $\sim 10$ years of observation with
the present instrumentation and collection rate of TOAs.

The still unassessed effects of Shapiro delay may also affect the new
determination of the binary eccentricity, for which in Paper\,I only
an upper limit was available.  Neglecting Shapiro delay, the measured
value is $e\,=\,3.4(12)\times10^{-6}$ (here and everywhere in this paper
the errors are quoted at twice the nominal rms values given by {\tt
TEMPO}). However, for $70\lapp i\lapp 89\degr$ and $0.17~{\rm
M_{\odot}}\lapp M_{co}\lapp 0.20~{\rm M_{\odot}}$ an unmodeled Shapiro
delay can introduce an apparent eccentricity in the range
$1-3\times10^{-6}$. The determination of $e$ must still be considered
provisional and $e\,=\,4.6\times10^{-6}$ is a reliable upper limit.

The small eccentricity of PSR\,A's binary system is typical of fully
recycled binary millisecond pulsars and is consistent with the effects
of random encounters with other cluster stars \citep{rh95}. The upper
limit on $e$ is also compatible with the offset position of PSR\,A
resulting from an interaction which occurred $\sim 1$ Gyr ago between the
already recycled binary system including PSR\,A and a WD companion
[PSR~A+WD] with a binary black hole of a few tens of solar masses
(\citealt{cmp03}, see\,\S\,1). We note that the value of $e$ also fits
with the hypothesis \citep{bvkh03,cmp03} that a
dynamical encounter with a single BH, whose mass is higher than a few
hundred ${\rm M_\odot}$, may have simultaneously ejected the
progenitor of [PSR~A+WD] and triggered the recycling process in the
binary, which in turn circularized the system and removed any
information on its post-encounter eccentricity. However, the
value of $e$ does not agree with an ejection event involving the
already formed [PSR\,A\,+\,WD] binary and a single BH. In this
case, \citet{cmp03} showed that the post-encounter eccentricity of
[PSR\,A+WD] would be significantly larger, up to values of
$10^{-4}-10^{-2}$ and only slightly affected by subsequent random
encounters with normal stars of the cluster \citep{rh95}.

The mean flux densities at 1400\,MHz ($S_{1400}$) in
Table~\ref{TabTim} are average values, derived from the system
sensitivity, the observed signal-to-noise ratio, the shape of the
pulse profile, the displacement of the pulsars with respect to the
center of the telescope beam and assuming flux density values
corresponding to half the detection limit for the non-detections due
to the strong interstellar scintillation effects on the pulsars in
NGC~6752 (see \S\,3). The uncertainties on the values of $S_{1400}$
may reach $\sim 30\%$ for the faintest sources. For a distance
$d=4.45\pm0.15$\,kpc \citep{gbc+03} the inferred radio luminosities at
1400 MHz of the two millisecond pulsars in the cluster's outskirts are
$L_{1400}\,=\,S_{1400}d^2\,\sim\,4-5$\,mJy\,kpc$^2$, a value in the
middle of the distribution of the luminosities of the millisecond
pulsars in 47~Tucanae \citep{clf+00}.
  
\section{Proper motion determinations}

The main improvement in our timing solutions is that proper motion
determinations for the two outermost pulsars in NGC\,6752 are now
available. In Table \ref{TabTim} proper motion components in right
ascension and declination are reported as well as the corresponding
proper motion amplitude and position angle (PA, measured
counterclockwise from north toward east). Proper motion uncertainties
depend on the length of the data span and on the number, the degree of
uniformity and the errors of the TOAs along the data span. The
different precisions in our measurements are mainly due to the
different number of high quality TOAs available for each pulsar, as
shown in Table \ref{TabTim}. Measurement of good TOAs for the
faintest pulsars is possible only when interstellar scintillation
enhances their signal: this is the reason that in the timing analysis
of PSRs B, D, and E we used a significantly smaller number of TOAs than for
PSR\,A. The flux density of PSR\,C is similar to that of PSR\,A and
the effects of interstellar scintillation are also comparable. The
difference in rms residual between the timing solutions for these two
pulsars is primarily due to the different pulse widths, being
$\sim\,7$ times larger (at 50\% of the peak) for PSR\,C than for
PSR\,A.

Figure\,\ref{fig:pm-map} presents a geometrical representation of the
expected motion in the plane of the sky (during next $10^4$\,yrs) of
PSR\,A, PSR\,C, and of the center of NGC\,6752, as derived from their
measured proper motions. The proper motion for the center of the
globular cluster has been obtained by \citet{dga99}, by comparing two
optical observations taken 25 years apart. The values
for the components are
$\mu_{\alpha}\cos\delta\,=\,-0.7\pm0.8$\,mas\,yr$^{-1}$ and
$\mu_{\delta}\,=\,-2.9\pm0.9$\,mas\,yr$^{-1}$. Their
derivation required a transformation of the coordinate system at the
epoch of the first observation to the coordinate system at the epoch
of the second observation and the use of distant field galaxies as
reference. The inset in Figure\,\ref{fig:pm-map} shows a comparison
between the proper motion vectors of PSR\,A and PSR\,C (with their
uncertainties), and the optical proper motion vector of the
cluster. The proper motions of PSR\,A and PSR\,C are
compatible with each other, but they are not in agreement with the
optical proper motion of NGC\,6752, at 4$\sigma$ and 2$\sigma$
confidence levels respectively.

\section{Discussion}

Since the escape velocity from a globular cluster is usually
significantly lower than the typical transverse velocity of these
stellar systems, it is expected that the proper motion of a cluster
pulsar will largely reflect the overall motion of the cluster. For
NGC\,6752, the escape velocity from the central region is
$\sim$\,30\,km\,s$^{-1}$ \citep{cmp03} and the space velocity is
$\sim$\,62\,km\,s$^{-1}$ with respect to the Solar System barycenter,
based on the proper motion measurement by \citet{dga99} and the
distance derived from the distance modulus
\citep{gbc+03}. Observations over a much longer data span may reveal
the peculiar (orbital) motion of a pulsar in the cluster's
gravitational potential well.

Is it possible that the discrepancy between the proper motions of
PSR\,A and PSR\,C and the optical proper motion of NGC\,6752 (\S\,3)
could be an indication that the two pulsars are not associated to the
globular cluster?

In Paper I it was estimated that the probability\footnote{It is worth
nothing that this probability does not account for the similar values
of the dispersion measure of PSR\,A and PSR\,C. Given the uncertainty
in the Galactic electron layer, it is difficult to quantify the
probability for this coincidence \citep{bkkv06}. However, it certainly
further decreases the total probability for a chance superposition.}
for PSR\,A to be a Galactic field millisecond pulsar superposed by
chance to NGC\,6752 (at a distance of $6\farcm4$ off its center) is of
order $10^{-5}.$ The compatibility of the measured proper motions of
PSR\,A and PSR\,C reinforces the unlikeliness for these two MSP to be
Galactic field objects by chance superimposed to the globular cluster.

Assuming that both PSRs A and C are members of NGC\,6752, the
discrepancy between pulsar and GC proper motions, measured in the
radio and optical band respectively, may result from the different
methods used for determining the proper motions in the two spectral
bands.  In fact, similar discrepancies have already been noted for the
pulsars in 47\,Tucanae \citep{fcl+01,fck+03}, in M4 \citep{tacl99}
and, more recently, in M15 \citep{jcj+06}.

However, the discrepancies in these clusters may not easily be
ascribed to a common systematic effect affecting all the optical
measurements. The optical proper motion for 47\,Tucanae was directly
measured based on Hipparcos observations. The proper motion
determination for M4 \citep{ch93} was based on the determination of
its motion relative to a set of reference field stars whose proper
motion relative to the Sun has been in turn obtained by combining
their position in the Galaxy (through their parallax) to a dynamic
model for the nearby regions of the Galaxy where these reference stars
reside.  Finally, the proper motion for NGC\,6752 \citep{dga+97}
resulted from the comparison of two photographic plates taken 25 years
apart, using distant field galaxies as reference objects. In the case
of M15, four different optical determinations have been performed
\citep{gclo93,soh+96,obgt97,ch93}, three of which are incompatible
with the pulsar proper motions. These three non-matching measures were
respectively based on the comparison between photographic plates at
different epochs \citep{gclo93,soh+96} and the use of reference stars
from Hipparcos observations \citep{obgt97}. Only the measurement by
\citet{ch93} is in agreement with the apparent motions of the three
pulsars investigated. However it is worth noting that \citet{ch93}
measured the optical proper motion for M15 applying the same method as
was used for M4, which in that case led to a discrepant proper
motion.

\begin{figure}
\centering
\plotone{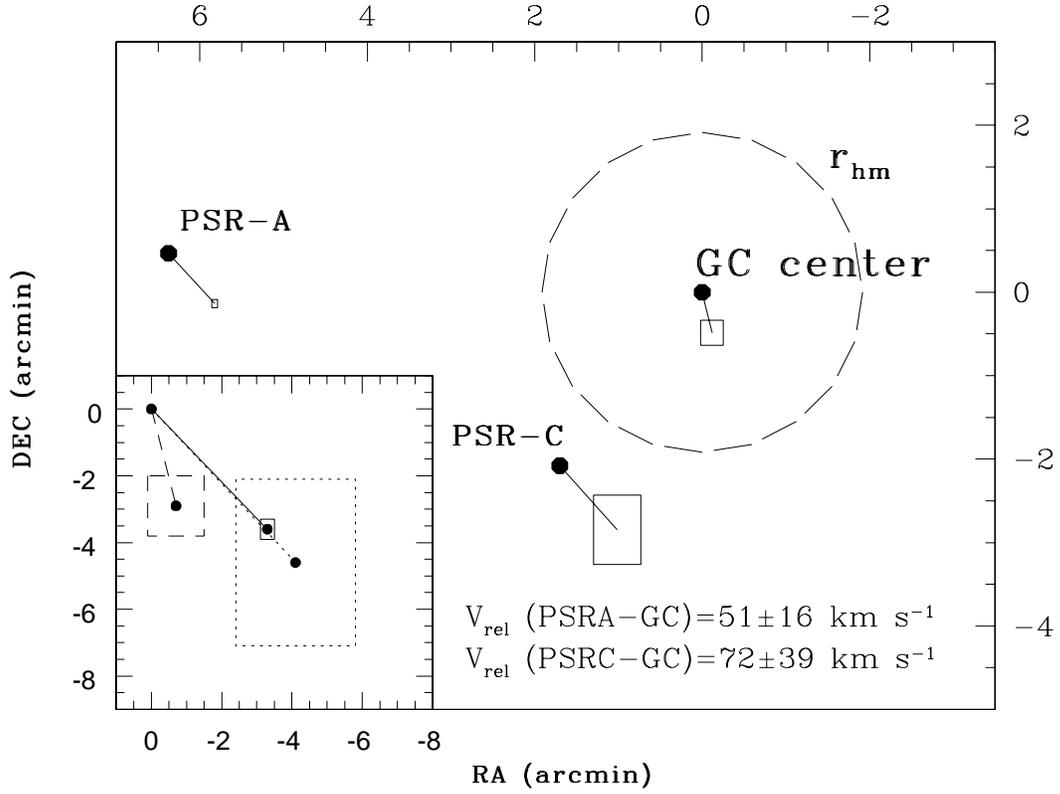}
\caption{{\it Main panel} - Positions and expected changes
(assuming uniform motion) after 10$^{4}$ years for PSR\,A, PSR\,C and
the center of NGC\,6752 relative to the present position of
the cluster center. The uncertainties in the expected
final positions are described by boxes whose size is given by the
propagation of the uncertainties on the proper motions in right
ascension and declination (2$\sigma$ confidence level).  Proper motion
uncertainties for the pulsars are from Table\,\ref{TabTim}, while the
uncertainties for the optical proper motion of the cluster are from
\citet{dga99}. The dashed circle represents the portion of the cluster
enclosed within the half-mass radius $r_{\rm hm}=1\farcm9$
\citep{tdk93}.
{\it Lower-left panel} - Comparison of the motions of the two
outermost pulsars (PSR\,A: solid line; PSR\,C: dotted line)and the
globular cluster (dashed line) relative to their present position.}
\label{fig:pm-map}
\end{figure}

The discrepancy may be alternatively ascribed to
very fast peculiar motions of PSR\,A and PSR\,C inside the cluster
gravitational potential well. Assuming that both the pulsar proper
motions and the optical proper motion of the cluster are correct, the
relative 2-D velocity vectors of the pulsars with respect to the
cluster center are roughly directed towards the cluster inner regions,
as is indicated in Figure \ref{fig:pm-map}. This
would mean that [PSR\,A+WD] cannot be now in the phase of ejection
from the cluster and that it is not at its farthest distance
from the GC center along its orbit inside the cluster gravitational
well.  For $d\,=\,4.45\pm0.15$\,kpc \citep{gbc+03} the relative
transverse speed of PSR\,A would be $V_{\rm
rel,A}\,=\,51\pm16$\,km\,s$^{-1}$. NGC\,6752 can provide a
gravitational pull strong enough to retain PSR\,A at its actual
location with a peculiar velocity $V_{\rm rel,A}$ only if the mass
enclosed within the pulsar projected position is $M_{\rm
encl}\,\geq\,1.18\times10^{6}\,{\rm M_\odot}.$ This is in contrast to
the total mass value for the cluster obtained with HST observation
\citep{sfs+04}, which is lower by an order of magnitude. Using again
the distance modulus in \citet{gbc+03}, the apparent magnitude given
by \citet{har96} and the colour excess $E_{\rm B-V}\,=\,0.04$ in
\citet{fmf+99}, the resulting overall mass-to-light ratio would be
$M_{\rm encl}/L\,\geq\,8.4\,{\rm M_\odot}/{\rm L_\odot}$, which is
unreasonably high for a GC, unless we admit an initial mass function
much flatter than usually estimated (so producing a very large number
of under-luminous stellar remnants) or the presence in the cluster of
a significant amount of dark matter. At a more conservative confidence
level (4$\sigma$) for the relative velocity $V_{\rm rel,A},$ the mass
to light ratio would result $M_{\rm encl}/L\,\geq\,2.5\,{\rm
M_\odot}/{\rm L_\odot}$, again implying a dynamic mass much higher
than the mass derived from optical observations.

A further test of the cluster membership of PSR\,A and
PSR\,C will be possible in future. It will involve the comparison
of the proper motions of PSR\,A and PSR\,C with those of the three
pulsars close to the cluster core, whose association with NGC\,6752 is
unambigously proved by the very strong gravitational pull affecting
the values of their spin period derivative. This task will take some
years: our simulations show that, with the present accuracy and
collection rate of the TOAs and if the three innermost pulsars display
the same proper motion as PSR\,A, a $3\sigma$ determination will
require a total data span of about $8-10$ years.

\acknowledgements
{\small AC, AP and NDA acknowledge the financial support to this
reasearch provided by the {\it Ministero dell'Istruzione,
dell'Universit\`a e della Ricerca} (MIUR) under the national program
{\it PRIN05 2005024090\_002}. The Parkes radio telescope is part of
the Australia Telescope which is funded by the Commonwealth of
Australia for operation as a National Facility managed by CSIRO.}


\newpage 

\newpage 

\newpage 

\newpage
\clearpage

\begin{deluxetable*}{llllll}
\tabletypesize{\footnotesize}
\tablecaption{Measured and derived parameters for the pulsars in NGC\,6752.\label{TabTim}}
\tablewidth{0pt}
\tablehead{
\colhead{Parameter} & \colhead{PSR\,A} & \colhead{PSR\,B} &
\colhead{PSR\,C} & \colhead{PSR\,D} & \colhead{PSR\,E}
}
\startdata
R.A.~(J2000) & 19:11:42.75562(8) & 19:10:52.0556(5) & 19:11:05.5552(4) & 19:10:52.4163(5) & 19:10:52.1572(6)\\
Decl.~(J2000)&--59:58:26.904(1) & --59:59:00.861(6) &
--60:00:59.700(4) & --59:59:05.479(5) & --59:59:02.087(7)\\
$\mu_{\alpha}\cos\delta$ (mas yr$^{-1}$) & --3.3(2) & -- & --4.1(17) & -- & --\\
$\mu_{\delta}$ (mas yr$^{-1}$) & --3.6(3) & -- & --4.6(25) & -- & --\\
$\mu$ (mas yr$^{-1}$) & 4.8(3)&--& 6.2(22)&--& --\\
PA$^{a}$(deg) &222(3)&--&221(20)&--& --\\
$P$~(ms) & 3.2661865707908(1) & 8.357798500844(2) &
5.2773269323093(15) & 9.035285247765(4) & 4.571765939750(2)\\
$\dot{P}^{b}~$(s\,s$^{-1}$) & 2.947(2)$\times10^{-21}$ & --7.9041(5)$\times10^{-19}$ & 2.16(2)$\times10^{-21}$
& 9.6431(6)$\times10^{-19}$ & --4.3435(3)$\times10^{-19}$\\
Epoch (MJD) & 51920.0000 & 52000.0000 & 51910.0000 & 51910.0000 & 51910.0000\\
DM~(pc~cm$^{-3}$) & 33.705(3) & 33.33(6) & 33.29(5) & 33.28(2) & 33.31(3)\\
$P_{orb}$~(days)       & 0.8371134769(1) & -- & -- & -- & -- \\
$a\sin i$~(l-s)  & 1.2060461(8) & -- & -- & -- & -- \\
$T_{asc}$~(MJD) & 51919.20647998(16) & -- & -- & -- &--\\
$e\sin\omega$   & 3.3(12)$\times10^{-6}$ & -- & -- & -- & -- \\
$e\cos\omega$   & 0.9(13)$\times10^{-6}$ & -- & -- & -- & -- \\
$f\left(M_{\rm c}\right)$ (${\rm M_\odot}$) & 0.002687854(6)& -- & -- & -- & -- \\
M$_{c,min}$$^{c}$\,(${\rm M_\odot}$) & 0.19 & -- & -- & -- & -- \\
MJD Range           & 51710--53836 & 51745--53769 & 51710--53836 & 51745--53731 & 51744--53836\\
Number~of~TOAs                    & 450  & 44   & 246  & 124  & 70    \\
r.m.s. residuals~($\mu$s)            & 5.0  & 18 & 29 & 24 & 25  \\
Offset$^{d}$~(arcmin)             & 6.37 & 0.06 & 2.56 & 0.05 & 0.05  \\
$W_{10}$$^{e}$ @10\% (ms) & 0.6 & 1.3 & 2.8 & 1.1 & 1.1  \\
$W_{50}$$^{f}$ @50\% (ms)   & 0.4 & 0.6 & 1.3 & 0.7 & 0.6  \\
$S_{1400}$ (mJy)   & 0.21 & 0.05 & 0.24 & 0.05 & 0.07  \\
\enddata
\tablenotetext{a}{Position angle of the proper motion vector.}
\tablenotetext{b}{As discussed in Paper I, the observed $\dot{P}$ of
PSR\,B, PSR\,C, PSR\,D, and PSR\,E are strongly affected by the
gravitational potential well of the globular cluster. Useful
constraints on the instrinsic spin-down rate $\dot{P}_{i}$ can hence
be inferred only for PSR\,A.  Correcting the observed value of
$\dot{P}$ for {\it (i)} the Galactic differential rotation and the
vertical acceleration in the Galactic potential (see
e.g. \citealt{dt91}), for {\it (ii)} the centrifugal acceleration of
the pulsar \citep{shk70}, and for {\it (iii)} the contribution of the
cluster potential well (estimated according to the recipe of
\citet{phi92} and using the luminosity density profile of NGC\,6752
published by \citet{fps+03}) gives $\dot{P}_{i}\lapp 6\times10^{-21}$
s s$^{-1}$. We have also adopted $M/L_{V}=5.5$ \citep{fps+03} in
order to obtain the firmest upper limit on the intrinsic spin-down
rate of PSR\,A.  This translates to a lower limit to the pulsar
spin-down age of $0.5(P/\dot{P}_{i})\sim 8.6$ Gyr and upper limits to
the surface dipole magnetic field of $3.2\times
10^{19}\sqrt{P\dot{P}_{i}}\sim 1.4\times 10^8$ G and to the spin-down
luminosity of $4\pi^2I\dot{P}_{i}/P^3=6.9\times 10^{33}$ erg s$^{-1}$
($I$ being the moment of inertia of the neutron star, set equal to
$10^{45}$ g cm$^{2}$).}  
\tablenotetext{c}{The minimum mass is
calculated assuming a pulsar mass of 1.35 ${\rm M_\odot}$ and an
inclination for the orbital plane with respect to the line of sight of
$90^{\circ}.$} 
\tablenotetext{d}{The offset of the pulsars is
calculated with respect to the position of the cluster's center of
gravity reported by \citet{fps+03}.}  
\tablenotetext{e}{Pulse width at 10\% of
the height of the main peak.}  
\tablenotetext{f}{Pulse width at 50\%
of the height of the main peak.}
\end{deluxetable*}

\end{document}